\documentclass[preprint,floatfix] {revtex4} 
 
\usepackage{graphicx}
\usepackage{subfigure}
\usepackage{amsmath}
\usepackage{xcolor}
\usepackage{color, soul}
\begin{document}

\title{Critical parameters and spherical confinement of H atom in screened Coulomb potential}
\author{Amlan K. Roy}
\altaffiliation{Email: akroy@iiserkol.ac.in, akroy6k@gmail.com, Ph: +91-3473-279137, 
Fax: +91-33-25873020.}
\affiliation{Division of Chemical Sciences, \\   
Indian Institute of Science Education and Research Kolkata, 
Mohanpur Campus, Nadia, 741246, India}

\begin{abstract}

\vspace{0.5in}
\noindent

Critical parameters in three screened potentials, namely, Hulth\'en, Yukawa and exponential cosine screened 
Coulomb potential are reported. Accurate estimates of these parameters are given for each of these potentials,
for all states having $n \leq 10$. Comparison with literature results is made, wherever possible. Present 
values compare excellently with reference values; for higher $n,\ell$, our results are slightly better. Some 
of these are presented for first time. Further,
we investigate the spherical confinement of H atom embedded in a dense plasma modeled by an exponential 
cosine screened potential. Accurate energies along with their variation with respect to box size and 
screening parameter are calculated and compared with reference results in literature. Sample
dipole polarizabilities are also provided in this case. The generalized pseudospectral method is used for 
accurate determination of eigenvalues and eigenfunctions for all calculations.  \\

{\bf\emph{Keywords:}} Screened Coulomb potential, Hulth\'en potential, Yukawa potential, critical 
screening, spherical confinement, generalized pseudospectral method. 
\end{abstract}
\maketitle

\section{Introduction}
Screened Coulomb potentials, $V(r)=-\frac{Z}{r} \sum_{k=0}^{\infty} V_k ({\color{blue}\delta} r)^k$, play a 
significant role in microscopic fields. They are often used as approximations to a number of 
interaction potentials in physics and chemistry, including atomic, molecular physics and quantum 
chemistry. When used in connection with atomic systems, $Z$ refers to atomic number, 
while the screening parameter {\color{blue} $\delta$} relates to different things in different branches. An 
enormous amount of work has been done on various aspects of these interacting potentials spanning 
nearly six decades. 

Our current communication focuses on three screened potentials, namely Hulth\'en, Yukawa and 
exponential cosine screened Coulomb (ECSC) potential. First one is an important 
short-range potential with relevance in atomic, solid-state and chemical physics; it is a special 
case of Eckart potential. In smaller $r$ region, both Hulth\'en and Yukawa potentials resemble 
Coulomb potential, while they decay monotonically exponentially to zero in larger $r$ region.  
Other than the $\ell=0$ states of Hulth\'en potential \cite{flugge74}, \emph{exact} analytical 
results remain unavailable for any of these systems; this has inspired a vast amount of 
publications for their bound and continuum states. Literature is quite extensive and we cite here 
some selected works. Bound states of Hulth\'en potentials are studied by variational 
\cite{varshni90, stubbins93}, shifted $1/N$ expansion \cite{tang87}, perturbation 
\cite{matthys88}, generalized pseudospectral (GPS) \cite{roy05}, asymptotic iteration 
\cite{bayrak07}, factorization \cite{dong07}, Nikiforov-Uvarov \cite{ikhdair07}, supersymmetry 
\cite{gonul00, qian02}, numerical \cite{nunez93} method, an algebraic approach \cite{setare07},
as well as numerous approximation schemes for centrifugal term \cite{jia09,ikhdair09,ikhdair11}, 
Laguerre pseudospectral method \cite{alici15}, etc. Likewise, bound states of Yukawa potential 
were also investigated by a host of approaches such as, combined Pad\'e approximation and 
perturbation theory \cite{vrscay86}, variational \cite{gomes94}, asymptotic iteration method 
\cite{karakoc06}, within the frame of Riccati equation \cite{gonul06}, tridiagonal 
matrix approach through a suitable Laguerre basis \cite{bahlouli10}, generalized parametric 
Nikiforov-Uvarov \cite{hamzavi12}, a direct method \cite{pena15} within Green-Aldrich 
approximation for centrifugal term, etc. 

Recently, there has been a surge of interest in the ECSC potential in solid-state, nuclear, 
plasma physics and field theory. Some of these are: one-electron atoms in various different 
shielding environments \cite{lin10a, lin10b,soylu12, chang13, lumb14}, ground and excited resonances 
of He, $H^{-}$ \cite{kar06, kar07, ghoshal09a, ghoshal09b}, molecular H$_2^+$ \cite{ghoshal11} in dense 
plasma using highly correlated wave function, two-electrons embedded in plasma within the 
configuration interaction framework \cite{ancarani14}, bound-state energies, polarizabilities, 
oscillator strengths of He \cite{lin15}, and etc. Since exact analytical results are not available 
for this potential, an impressive amount of work has been reported for their eigenspectra. For 
example, variation and perturbation method \cite{lam72}, and Ecker-Weizel \cite{dutt79,ray80} and
hypervirial-Pad\'e \cite{lai82} approximation, a dynamical group approach \cite{meyer85}, 
hypervirial equation with Hellman-Feynman theorem \cite{sever90}, a numerical method 
\cite{singh83}, large-N expansion \cite{sever87}, shifted 1/N expansion \cite{ikhdair93}, asymptotic 
iteration \cite{bayrak07a}, perturbation \cite{ikhdair07a}, J-matrix \cite{nasser11}, Ritz 
variation \cite{paul11}, GPS \cite{roy13}, Laguerre pseudospectral \cite{alici15} method etc.  

A distinctive feature that characterizes screened potentials is the presence of a \emph{limited} 
number of bound states (in contrast to Coulomb potential). For each $(n, \ell)$ eigenstate, there 
exists a certain threshold value of screening parameter at which the binding energy of a given level 
in question becomes zero. That means, beyond this \emph{critical} screening parameter 
($\delta=\delta_c$), no bound states could be found, so that at this point, $E(\delta_c)=0$. While 
one can find a huge amount of reference works for bound and resonant states of these potentials 
(as mentioned earlier), same for critical screening is rather scarce. Nevertheless, a few results
are available in the literature, which are cited herein. These are published for Hulth\'en potential
in \cite{popov85, varshni90, demiralp05}, Yukawa potential in \cite{rogers70, diaz91, gomes94} and ECSC potential in 
\cite{lam72, ray80, singh83, diaz91, nasser11}. {\color{red} Very recently, the critical parameters of
$1sns$ $^{1,3}S^e$ and $1snp$ $^{1,3}P^o$ ($n \leq 5$) states of He immersed in weakly coupled Debye plasma, 
modeled by screened Yukawa potential, have been studied \cite{lin15a} as well.} A primary objective of this work is to 
report accurate estimates of critical parameters for all three above mentioned potentials. To this end, all 
55 states corresponding to $n \leq 10$ are considered systematically. For this, we employ the GPS 
method which has been demonstrated to produce quite reliable and accurate results for a variety of 
physically and chemically important systems including quantum confinement \cite{roy04, roy04a, roy05, 
roy08, roy08a, roy13, roy15}. 

A secondary objective of this work is to investigate spherical confinement of H atom embedded in 
dense plasma modelled by ECSC potential. This takes inspiration from a recent publication 
\cite{lumb14}, where interaction 
of such a system with short laser pulses in femtosecond regime was studied recording the 
effects of confinement radius, Debye screening length as well as laser parameters  such as 
strength, shape, frequency and duration of pulse. Quantum confinement of H atom and other central
potentials within an impenetrable spherical cavity has been a subject of much current interest. Many 
interesting, fascinating phenomena occur under such small spatial dimensions relative to the 
corresponding \emph{free} systems. Literature on the topic is vast and rich; interested reader 
may consult the special issues in \emph{Advances in Quantum Chemistry} \cite{sabin09} as well as 
the recent book \cite{sen14} and numerous references therein. A detailed analysis of the energy 
spectrum is presented here considering the effect of confinement radius and screening parameter 
on energy levels. Both low and high-lying states are treated for \emph{small, medium and large} $r_c$. 
Additionally some specimen dipole polarizabilities are given as well. Giving a brief account of 
the method in Sec.~II, we proceed for results in Sec.~III. A few remarks are made in Conclusion 
in Sec.~IV. 

\section{Method of calculation}
The GPS method has been shown to be a simple useful and powerful approach for a variety of
physical systems. They were discussed previously in a number of communications \cite{roy04, 
roy04a, roy05, roy08, roy08a, roy13, roy15}; hence not repeated here. Spherical confinement 
of a particle in a central potential is modeled, without any loss of generality, by the following 
radial non-relativistic Schr\"odinger equation (atomic unit employed unless otherwise mentioned):
\begin{equation}
H \psi_{n,\ell} (r) = 
\left[ -\frac{1}{2} \frac{\mathrm{d^2}}{\mathrm{dr^2}} + \frac{\ell (\ell+1)}{2r^2} + 
v(r)+v_c(r) \right] \psi_{n,\ell} (r), 
\end{equation}
where $n,\ell$ signify usual radial and angular quantum numbers, while $v(r)$ characterizes 
the particular screening potential under consideration. For corresponding \emph{free} systems, 
$v_c(r)=0$, whereas confinement is achieved by the following equation ($r_c$ denotes confining 
radius), 
\begin{equation} v_c(r) = \begin{cases}
+\infty \ \ \ \ r> r_c  \\
0,  \ \ \ \ \ \ \ r \leq r_c.   \\
 \end{cases} 
\end{equation}
Here we are interested in the following three cases, 
\begin{equation} v(r)= \begin{cases} 
-\frac{\delta e^{-\delta r}}{1-e^{-\delta r}}  \ \ \ \ \ \ \ \ \ \  \ \ \ \ \mathrm{Hulth\acute{e}n} \ 
\mathrm{potential} \\
-\frac{e^{-\delta r}}{r}  \ \ \ \ \ \ \ \ \ \ \ \ \ \  \ \ \  \mathrm{Yukawa} \ \mathrm{potential} \\ 
-\frac{e^{-\delta r}}{r}  \cos (\delta r) \ \ \ \ \ \  \ \mathrm{ECSC} \ \mathrm{potential}. 
\end{cases}
\end{equation} 
For convenience, same screening parameter $\delta$ is used for all three. Eigenvalues, eigenfunctions 
are obtained by solving Eq.~(1) satisfying the boundary condition 
{\color{red}
$\psi_{n, \ell}\ (0) \  \mathrm{finite} \  \mathrm{and} \  \psi_{n,\ell} \ (r_c)=0$.}

A key feature of this method is that a given function defined in the semi-infinte domain 
$r \in [0,\infty]$ is approximated by an $N$-th order polynomial in finite interval $[-1,1]$, such 
that at the \emph{collocation points}, approximation is \emph{exact}. This facilitates working in a 
\emph{non-uniform, optimal} spatial discretization, where a relatively smaller number of radial point 
leads to sufficiently good accuracy. Through a non-linear mapping and a symmetrization procedure, 
this generates a finer mesh at smaller $r$ and cruder mesh at larger $r$, preserving similar kind of 
accuracy in both these regions. Eventually this leads to \emph{symmetrical eigenvalue} problem, which 
can be easily solved accurately by means of standard routines {\color{red} (from NAG libraries) available. 
Energy calculations were performed with successive incremental changes in $\delta$; critical parameters 
were recorded by noting a change in the sign of energy values.}

\begingroup
\squeezetable
\begin{table}
\caption {\label{tab:table1} Estimated critical screening parameters of Hulth\'en, Yukawa and ECSC 
potential for some low-lying states having $n=1-5, \ell=0-4$. See text for details.} 
\begin{ruledtabular}
\begin{tabular}{lllllll}
 State & \multicolumn{2}{c}{$\delta_c$ (Hulth\'en)} & \multicolumn{2}{c}{$\delta_c$ (Yukawa)} &   
 \multicolumn{2}{c}{$\delta_c$ (ECSC) }   \\ 
\cline{2-3} \cline{4-5} \cline{6-7}
  &  PR$^\dag$   & Ref.         &   PR$^\dag$   &    Ref.       &  PR$^\dag$   &     Ref.     \\    \hline
 $1s$   &  2.00000000  & 2.000000\footnotemark[1]    
        &  1.190610    & 1.190612\footnotemark[2],
        & 0.7205240    & 0.7115\footnotemark[4],0.7131\footnotemark[5],0.72052408\footnotemark[6],  \\
        &    &    &    & 1.190612\footnotemark[3]   &     &   0.72055425\footnotemark[7]      \\             

 $2s$   &  0.49999999  & 0.500000\footnotemark[1]    
        &  0.310199    & 0.310209\footnotemark[3]                                         
        & 0.1666172    & 0.18605\footnotemark[5],0.1666176\footnotemark[6],0.16656630\footnotemark[7]  \\

 $3s$   &  0.22222222  & 0.222222\footnotemark[1]    
        &  0.139466    & 0.139450\footnotemark[3]                                        
        & 0.07243689   & 0.08828\footnotemark[5],0.07243699\footnotemark[6],0.07242453\footnotemark[7] \\

 $4s$   &  0.1249999   & 0.125000\footnotemark[1]    
        &  0.078825    & 0.078828\footnotemark[3]                                      
        &  0.04042716  & 0.0513\footnotemark[5],0.04042722\footnotemark[6],0.04042424\footnotemark[7]  \\

 $5s$   &  0.0799999   & 0.080000\footnotemark[1]    
        &  0.050580    & 0.050583\footnotemark[3]                                      
        & 0.02578729   & 0.02578730\footnotemark[6],0.02578635\footnotemark[7]   \\

 $2p$   & 0.37693599   & 0.376936\footnotemark[1],
        & 0.220216806  & 0.220216806\footnotemark[3]         
        & 0.1482050325 & 0.143\footnotemark[4],0.14820503\footnotemark[6],0.148205032\footnotemark[7]   \\
        &              & 0.376759\footnotemark[8]  &   &    &    &         \\        

 $3p$   & 0.18648588   & 0.186486\footnotemark[1],             
        & 0.112710498  & 0.112710498\footnotemark[3]              
        & 0.0687121435 & 0.066\footnotemark[4],0.06871214\footnotemark[6],0.068712143\footnotemark[7]   \\
        &              & 0.186364\footnotemark[8]  &   &    &    &         \\ 

 $4p$   & 0.1104912    & 0.110491\footnotemark[1],
        & 0.067885376  & 0.067885376\footnotemark[3]        
        & 0.0392634011 & 0.037\footnotemark[4],0.03926340\footnotemark[6],0.039263401\footnotemark[7]   \\
        &              & 0.110410\footnotemark[8]  &   &    &    &         \\ 

 $5p$   & 0.0728634    & 0.072863\footnotemark[1],
        & 0.045186248  & 0.045186248\footnotemark[3]    
        & 0.0253156252 & 0.024\footnotemark[4],0.02531562\footnotemark[6],0.0253156248\footnotemark[7]  \\
        &              & 0.072806\footnotemark[8]  &   &    &    &         \\ 

 $3d$   &  0.15766196  & 0.157662\footnotemark[1],
        &  0.091345120 & 0.091345120\footnotemark[3]     
        & 0.0635815461 & 0.062\footnotemark[4],0.06358154\footnotemark[6],0.063581546\footnotemark[7]  \\
        &              & 0.157659\footnotemark[8]  &   &    &    &         \\ 

 $4d$   &  0.09756383  & 0.097564\footnotemark[1], 
        &  0.058105052 & 0.058105052\footnotemark[3]         
        & 0.0374050483 & 0.036\footnotemark[4],0.03740505\footnotemark[6],0.037405048\footnotemark[7]  \\
        &              & 0.0975606\footnotemark[8]  &   &    &    &        \\

 $5d$   &  0.06610780  & 0.066108\footnotemark[1],
        &  0.040024353 & 0.040024353\footnotemark[3]             
        & 0.0245000141 & 0.024\footnotemark[4],0.02450001\footnotemark[6],0.024500014\footnotemark[7]  \\
        &              & 0.0661043\footnotemark[8]  &   &    &    &        \\  

 $4f$   &  0.08640509  & 0.086405\footnotemark[1],
        &  0.049831132 & 0.049831132\footnotemark[3]        
        & 0.0352412421 & 0.035\footnotemark[4],0.03524124\footnotemark[6],0.035241242\footnotemark[7] \\
        &              & 0.08640507\footnotemark[8] &   &    &    &        \\

 $5f$   &  0.05997272  & 0.059973\footnotemark[1],
        &  0.035389389 & 0.035389389\footnotemark[3]   
        & 0.0234821564 & 0.023\footnotemark[4],0.02348216\footnotemark[6],0.023482156\footnotemark[7] \\
        &              & 0.05997268\footnotemark[8] &   &    &    &        \\

 $5g$   & 0.05450531   & 0.054505\footnotemark[1],             
        & 0.031343552  &                                          
        & 0.0223714239 & 0.022\footnotemark[4],0.02237142\footnotemark[6],0.022371423\footnotemark[7] \\
        &              & 0.05450531\footnotemark[2]  &   &   &    &       \\
\end{tabular}
\end{ruledtabular}
\begin{tabbing}
$^{\mathrm{a}}$Ref.~\cite{varshni90}. \hspace{60pt} \= 
$^{\mathrm{b}}$Ref.~\cite{gomes94}. \hspace{55pt} \=
$^{\mathrm{c}}$Ref.~\cite{diaz91}. \hspace{55pt} \=
$^{\mathrm{d}}$Ref.~\cite{lam72}. \hspace{55pt} \= 
$^{\mathrm{e}}$Ref.~\cite{ray80}. \hspace{55pt}  \\
$^{\mathrm{f}}$Ref.~\cite{singh83}. \hspace{55pt} \=
$^{\mathrm{g}}$Ref.~\cite{nasser11}. \hspace{55pt} \=
$^{\mathrm{h}}$Ref.~\cite{demiralp05}. \hspace{55pt} \=
$^\dag$PR implies Present Result.
\end{tabbing}
\end{table}
\endgroup

\section{Results and Discussion}
\subsection{Critical screening}
{\color{red} At first, we discuss $\delta_c$ values for all $n \leq 5, \ell=0-4$ states of Hulth\'en potential
in columns 2, 3 of Table~I. The remaining 45 states corresponding to $6 < 5 \leq 10$ are presented in 
Table~S1 of the \emph{Supporting Document}.}
The presented results here and in following tables were thoroughly checked for convergence by running 
a series of calculations changing length of grid (see discussion below) and mapping parameter 
(see GPS references cited earlier for its definition); which we employ as 25. It is found that the 
results are insensitive towards variations in number of radial points (we used 200). As mentioned earlier, 
some literature results are available for all the states considered, which are duly quoted for comparison. 
First definitive attempt was made in \cite{popov85}, where reasonably good 
estimates were reported for $1s, 2p, 3d, 4f, 5g$ and $10m$ states by means of a perturbation theory 
summation method, nearly three decades ago. {\color{red} Thus, for above mentioned states, $\delta_c$'s of 
2.0000, 0.3768, 0.1577, 0.0864, 0.0545 and 0.0133 respectively, match up to three to four places in the 
decimal with present as well as other available literature values.} Later, 
these were revisited in the numerical calculation of \cite{varshni90}; $\delta_c$'s were systematically 
determined for all $n \leq 10$ states. Present results show decent agreement with these for all the 
states. About a decade ago, a variational approach \cite{demiralp05} based on a coordinate transformation 
on radial variable was also suggested for their calculation (for all $p-h$ states, i.e., $\ell=1-5; 
n=(\ell+1)$ to 10 for each $\ell$). In general, one notices good matching of GPS results with these two 
$\delta_c$s; probably ours are slightly better than the former two, especially for higher $n,\ell$ states.  
It is worth mentioning here that, as the screening parameters approached critical zone, considerably 
larger $R$ values were necessary; this fact was mentioned before in \cite{roy05}. Similar numerical 
instabilities have also occurred in \cite{demiralp05} as well, in the neighborhood of $\delta_c$. For 
$s$ states, $\delta_c$'s are readily obtained analytically as $\delta_c=2/n^2$ from the energy expression. 
Note that, a simple approximate analytic expression as the following,  
\cite{patil84}, 
{\color{red}
\begin{equation}
\delta_c=\frac{1}{\left( \frac{n} {\sqrt{2}}+0.1645\ell+0.0983 \ \frac{\ell}{n} \right)^2}.
\end{equation}} 
was also suggested for $\delta_c$'s in terms of quantum numbers $n, \ell$, which seem to be moderately 
good for whole range of $n,\ell$. To save space and avoid clumsiness, there are omitted here; interested 
reader may find them in \cite{varshni90}.

{\color{red} Next, our calculated $\delta_c$ values for Yukawa potential are tabulated in columns 4,5 of Table I, for same
15 states as in Hulth\'en potential; while the rest 40 states are given in Table~S2 of \emph{Supporting Document.}} 
Reference results are {\color{red} notably scarce} in this case. Several decades 
ago, these were first reported \cite{rogers70} via numerical integration of radial Schr\"odinger equation 
for states having $\ell=0-9; n=(\ell+1)$ to 9 for each $\ell$. Later, very precise 
(1.19061227$\pm$0.00000004 in a.u.) estimate of $\delta_c$ for ground state has been reported in a
variational calculation \cite{gomes94} through linear combination of atomic orbitals scheme. Our 
current approach does not offer such accuracy for ground state. Nevertheless, the agreement is quite 
reasonable noting that all these three methods differ only in last digit. {\color{red} A more systematic and accurate calculation
of these parameters for $n \leq 5; \ell =0-3$ states were performed in the propagation matrix solution 
\cite{diaz91}. For lowest $\ell$ states, some disagreement is observed between the GPS result and reference. 
However, for $\ell \geq 1$, the two results practically coincide with each other. For $n > 5$ states, no literature values
are available other than that of \cite{rogers70}. Our current estimates are significantly improved and we hope these 
could be useful for future referencing.} No results are available for $n=10$ states. As in Hulth\'en potential, here also, 
we had to enlarge radial coordinate considerably. Generally speaking, $\ell \neq 0$ states offer better accuracy 
than $s$ waves. 

Now we move on to $\delta_c$ in ECSC potential, in columns 6,7 of Table I. Variationally calculated $\delta_c$'s  
for ECSC potential were reported \cite{lam72} for ground state and those corresponding to $\ell=1-7; 
n=(\ell+1)$ to 8, with moderate accuracy. Also these for $ns$ states with $n=2-4$ were reported 
through an Ecker-Weizel method by approximating ECSC potential by a Hulth\'en potential 
\cite{ray80}. Later more accurate estimates for these parameters were published for all 36 
$n \leq 8$ states via numerical \cite{singh83} as well as J-matrix \cite{nasser11} methods. 
{\color{red} For $n \leq 6$, highly accurate critical parameters have been reported \cite{diaz91} 
through a propagation-matrix solution of the eigenvalue equation by means of some simple numerical scheme.} 
For $n > 8$, $\delta_c$'s remain unreported as yet. Some differences in our results with those from 
\cite{singh83} and \cite{nasser11} are recorded for $\ell=0$. Otherwise, present values are in good
accord with these two. Apparently those from \cite{nasser11} are slightly more precise than those
of \cite{singh83}. {\color{red} As in case of the previous two potentials, here also, the critical parameters 
for $5 < n \leq 10$ states are provided in Table~S3 of \emph{Supporting Document.}}

\begingroup
\squeezetable
\begin{table}
\caption {\label{tab:table2} Some low-lying states of confined ECSC potential, at selected $r_c$, for 
four $\delta$ values. Reference results are taken from \cite{lumb14}. PR implies Present Result. See
text for details.}
\begin{ruledtabular}
\begin{tabular}{l|llll|llll}
 $r_c$ &  E$_{1s}$(PR)   &   E$_{1s}$(Ref.)  &  E$_{2p}$(PR)   &   E$_{2p}$(Ref.)  &  E$_{1s}$(PR)  &   
          E$_{1s}$(Ref.)    &  E$_{2p}$(PR)    &   E$_{2p}$(Ref.)   \\   \hline
       & \multicolumn{4}{c}{$\delta=0.1$}  &  \multicolumn{4}{c}{$\delta=0.2$} \\
   \cline{2-9} 
0.1     &  469.093037729      &               &   991.107588200          &          
        &  469.193031240      &               &   991.207579548          &              \\
0.5     &  14.8479479912      &               &   36.7588456253          &         
        &  14.9477963784      &               &   36.8586378426          &             \\
1       &  2.47390897226      &               &   8.32302110803          &         
        &  2.57335504819      &               &   8.42223133406          &             \\
1.5     &  0.53684831143      &               &   3.33079617477          &   
        &  0.63571857088      &               &   3.42910868123          &             \\
2       & $-$0.02527539005    & $-$0.02528    &   1.67557999210          &  1.67558         
        &  0.07291957244      & 0.07292       &   1.77273282657          &  1.77273    \\
3       & $-$0.32446684128    &               &   0.58032931034          &  
        & $-$0.22765602515    &               &   0.67456878317          &             \\ 
5       & $-$0.39721777698    & $-$0.39722    &   0.10539295060          &  0.10539  
        & $-$0.30218038193    & $-$0.30218    &   0.19252014166          &  0.19252    \\
10      & $-$0.40088390676    & $-$0.40088    &  $-$0.02441874085        &
        & $-$0.30633267927    & $-$0.30633    &   0.04586711993          &  0.04587    \\
50      & $-$0.40088477463    & $-$0.40088    &  $-$0.03246880517        &  $-$0.03247
        & $-$0.30633448845    & $-$0.30633    &   0.00421458654          &  0.00421    \\
100     & $-$0.40088477464    &               &  $-$0.03246880518        &
        & $-$0.30633448845    &               &   0.00101563901          &             \\  
\hline
        & \multicolumn{4}{c}{$\delta=0.5$}  &  \multicolumn{4}{c}{$\delta=1$} \\
\hline
0.1     &  469.492923795      &               &  991.507436380           &
        &  469.992133802      &               &  992.006385267           &             \\
0.5     &  15.2453781208      &               &  37.1553387499           &         
        &  15.7287648772      &               &  37.6328672123           &             \\
1       &  2.86491282463      &               &  8.71032168771           &
        &  3.31149548661      &               &  9.13651977224           &             \\
1.5     &  0.91922572831      &               &  3.70493149798           &   
        &  1.32242862550      &               &  4.06841930273           &             \\
2       &  0.34758134919      & 0.34758       &  2.03394022194           &  2.03394
        &  0.70817929642      & 0.70818       &  2.33370405788           &  2.33370    \\
3       &  0.03025176422      &               &  0.90324729702           &               
        &  0.31880855543      &               &  1.09173878309           &             \\
5       &  $-$0.06497226478   & $-$0.06497    &  0.35727629336           &  0.35728      
        &  0.13351264273      & 0.13351       &  0.41381878384           &  0.41382    \\
10      &  $-$0.07749780230   & $-$0.07750    &  0.10554002869           &  0.10554     	
        &  0.04125952928      & 0.04126       &  0.10213224857           &  0.10213    \\
50      &  $-$0.07768368464   & $-$0.07768    &  0.00404455216           &  0.00404     	
        &  0.00191258278      & 0.00191       &  0.00403857403           &  0.00404    \\
100     &  $-$0.07768368464   &               &  0.00100973663           &      	
        &  0.00048577694      &               &  0.00100954749           &             \\
\end{tabular}
\end{ruledtabular}
\end{table}
\endgroup

\subsection{Spherically confined H atom in dense quantum plasma}
Now Table II offers energies of H atom confined at the center of an inert impenetrable cavity embedded
in an ECSC potential. Two lowest states ($1s, 2p$) corresponding to $\ell=0$, 1 are given at four 
different strengths (0.1, 0.2, 0.5 and 1) of screening parameter to cover \emph{weak, intermediate and 
strong} screening. Ten $r_c$'s have been chosen carefully in each case representing \emph{small, medium  
and large} range of confinement. As mentioned in Sec.~I, a substantial amount of work exists for the 
respective \emph{free} system. However, to the best of our knowledge, only one reported work 
\cite{lumb14} can be found for its confinement, where some eigenvalues (within the range 
$r_c \leq 2 \leq 50$) were given by employing a Bernstein polynomial approach. Wherever available, 
present GPS energies compare quite favorably with these, offering slightly better accuracy. Many new 
states are given here for the first time. 

Now, some representative moderately low-lying states of ECSC potential inside a spherical enclosure 
are offered in Table~S4 of \emph{Supporting Document}; energies are produced for all 7 states corresponding 
to $n=3,4$, for a fixed 
screening parameter, $\delta=0.02$. Ten selected $r_c$, \emph{viz.}, 0.1, 0.5, 1, 2, 5, 10, 20, 30, 
50 and 100 a.u., are chosen to scan the whole range of confinement. Similar to the confined H atom or 
Hulth\'en potential case \cite{roy15}, for a given $\delta_c$, starting from a large positive value at 
smaller $r_c$, energy steadily decreases monotonically, crossing zero and becoming negative at certain 
$r_c$, eventually approaching a constant value at a sufficiently large $r_c$ thereafter. 
Available reference energies are provided at four $r_c$, \emph{viz.,} 2, 5, 10 and 50 respectively,
which seem to match rather well with present energies. To the best of our knowledge, no further attempt 
is known for these states, and hopefully they may constitute a useful set of reference for future 
works in this direction. 

\begin{figure}         
\begin{minipage}[c]{0.55\textwidth}
\centering
\includegraphics[scale=0.48]{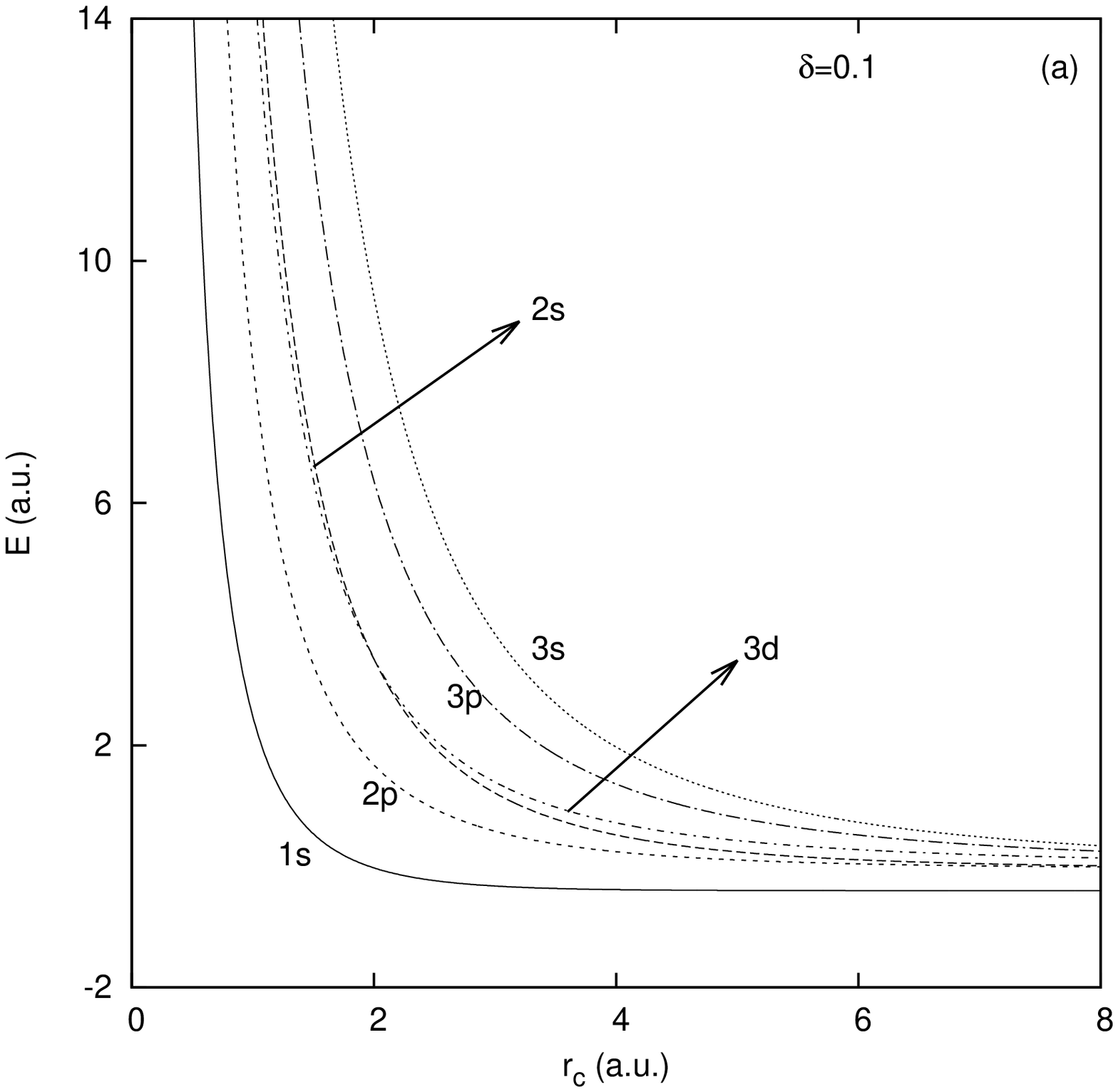}
\end{minipage}%
\begin{minipage}[c]{0.55\textwidth}
\centering
\includegraphics[scale=0.48]{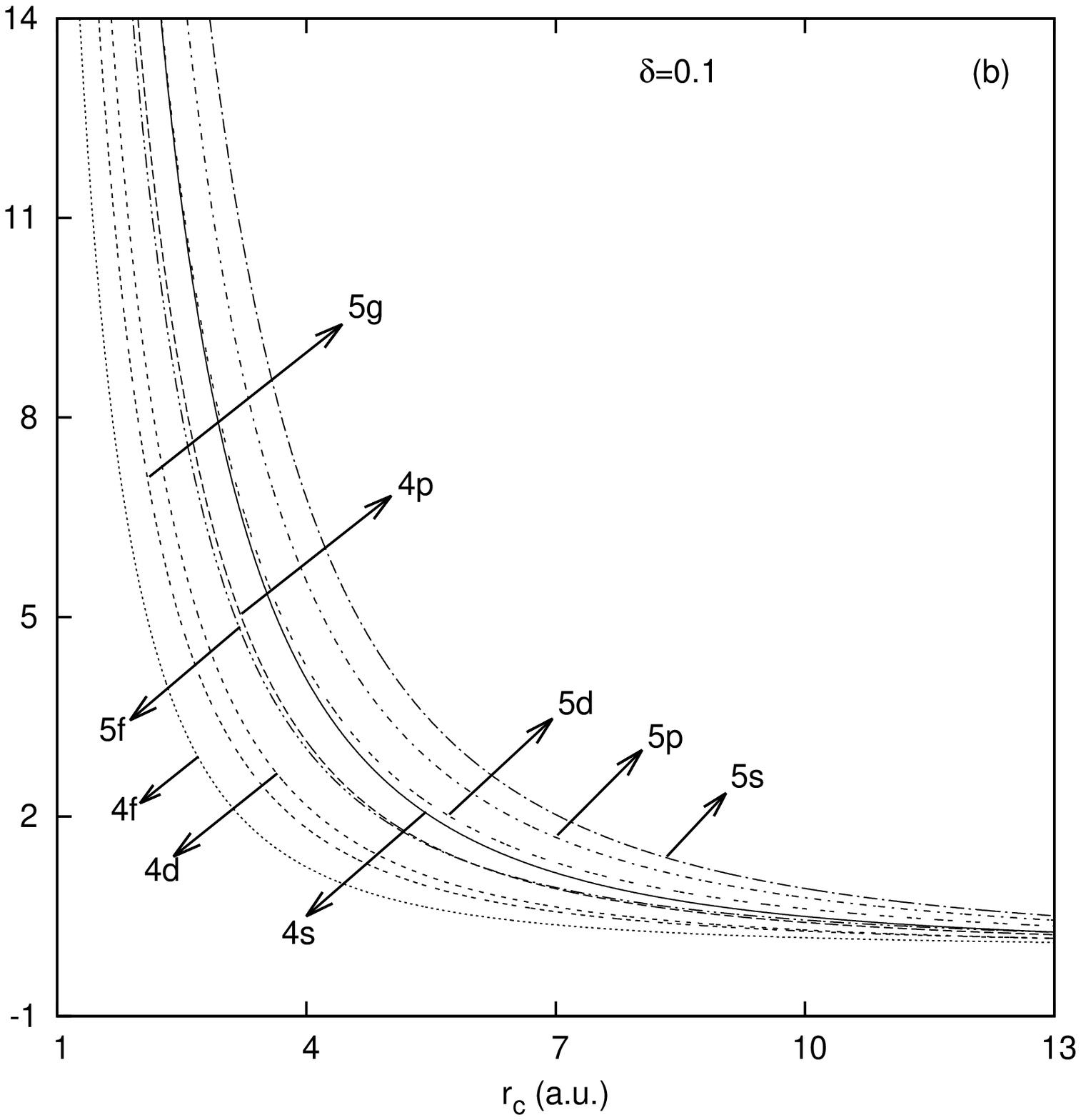}
\end{minipage}%
\caption[optional]{Energy variations in compressed ECSC potential, with respect to $r_c$, for $\delta=0.1$: 
(a) 6 states belonging to $n=1,2,3$; (b) 9 states corresponding to $n=4,5$.}
\end{figure}

Above variations of isotropic compression of energies of ECSC potential in Table~II are 
clearly depicted in energy vs. $r_c$ plots in Fig.~1, for a fixed $\delta=0.1$. In left panel (a), 
these are given for all six states corresponding to $n=1-3$; (b) shows similar plots for all nine 
states belonging to $n=4,5$. Both positive, negative energies are considered in all cases. Ranges 
of energy and $r_c$ axes differ from each other in these plots to 
clearly visualize the effects. Generally, shapes of these curves appear quite similar to each other; 
normally they also remain well separated (and parallel) for smaller $r_c$ merging at a sufficiently 
large $r_c$. Very small confinement is avoided for appreciation of figures. As $r_c$ is gradually 
reduced towards lower values, energies exhibit a sharp increase. Consistent with confinement in 
isotropic confinement in central potentials, starting from an initial high positive value, energies
tend to fall off rapidly monotonically with increase in $r_c$, finally approaching energy of 
corresponding free system smoothly and thereafter assumes a constant value. With decrease in $r_c$, 
energies change sign from negative to positive passing through a zero at critical cavity 
radius. One also notices crossing between some of these levels at certain $r_c$'s; in other words, these
states become degenerate at those respective $r_c$'s. Some such pairs are ($2s,3d$) in  
(a); ($4s,5d$), ($4p, 5f$) in (b), besides ($3p,4f$) (not shown), which is reminiscent of \emph{simultaneous 
degeneracy} encountered in confined H atom. Furthermore, it is found that, for a 
specific $\delta$, with decrease in $r_c$, states having same $\ell$ and different $n$ maintain 
separation; no mixing occurs among them. Thus, for a particular $\ell$ and $r_c$, state with lowest $n$ 
remains lowest in energy and vice versa, such that one finds the following energy ordering: 
$E_{1s} < E_{2s} < E_{3s} < E_{4s} \cdots$; $E_{2p} < E_{3p} < E_{4p} \cdots $; $E_{3d} < E_{4d} \cdots$, 
etc. Likewise, for a specific $\delta$, within a given $n$, individual $\ell$ levels remain 
well separated at a given $r_c$ without crossing each other, finally attaining the energy of respective 
\emph{free} system at a large $r_c$. With decrease in $r_c$, state with higher $\ell$ gets relatively
stabilized such that the state with largest $\ell$ becomes lowest in energy and vice versa. So for a 
given $n$, the orderings are found as: $E_{2p} < E_{2s}$; $E_{3d} < E_{3p} < E_{3s}$; 
$E_{4f} < E_{4d} < E_{4p} < E_{4s}$, etc. Similar energy orderings were found in confined H atom 
and Hulth\'en potential \cite{roy15}. As enclosure size is reduced, numerous complex energy splitting 
is observed, especially for states having high $n,\ell$ quantum numbers. Moreover, for a specific $\delta$, 
level ordering follows the \emph{same} pattern as H atom and Hulth\'en potential under similar spherical
confinement; in the limit of $r_c \rightarrow 0$ this gives, 
\[ 1s,2p,3d,2s,4f,3p,5g,4d,6h,3s,5f,7i,4p,8k,6g,5d,4s,9l,7h,6f,10m,5p,8i, \cdots \]
This has been checked for several $\delta$'s. However, as confining radius increases, energy ordering is 
characterized by frequent intermixing between levels belonging to different $n$ values. In intermediate 
and large $r_c$ region, different $\delta$ seems to provide different orderings. 

\begin{figure}     
\vspace{-1in}
\begin{minipage}[c]{0.55\textwidth}
\centering
\includegraphics[scale=0.48]{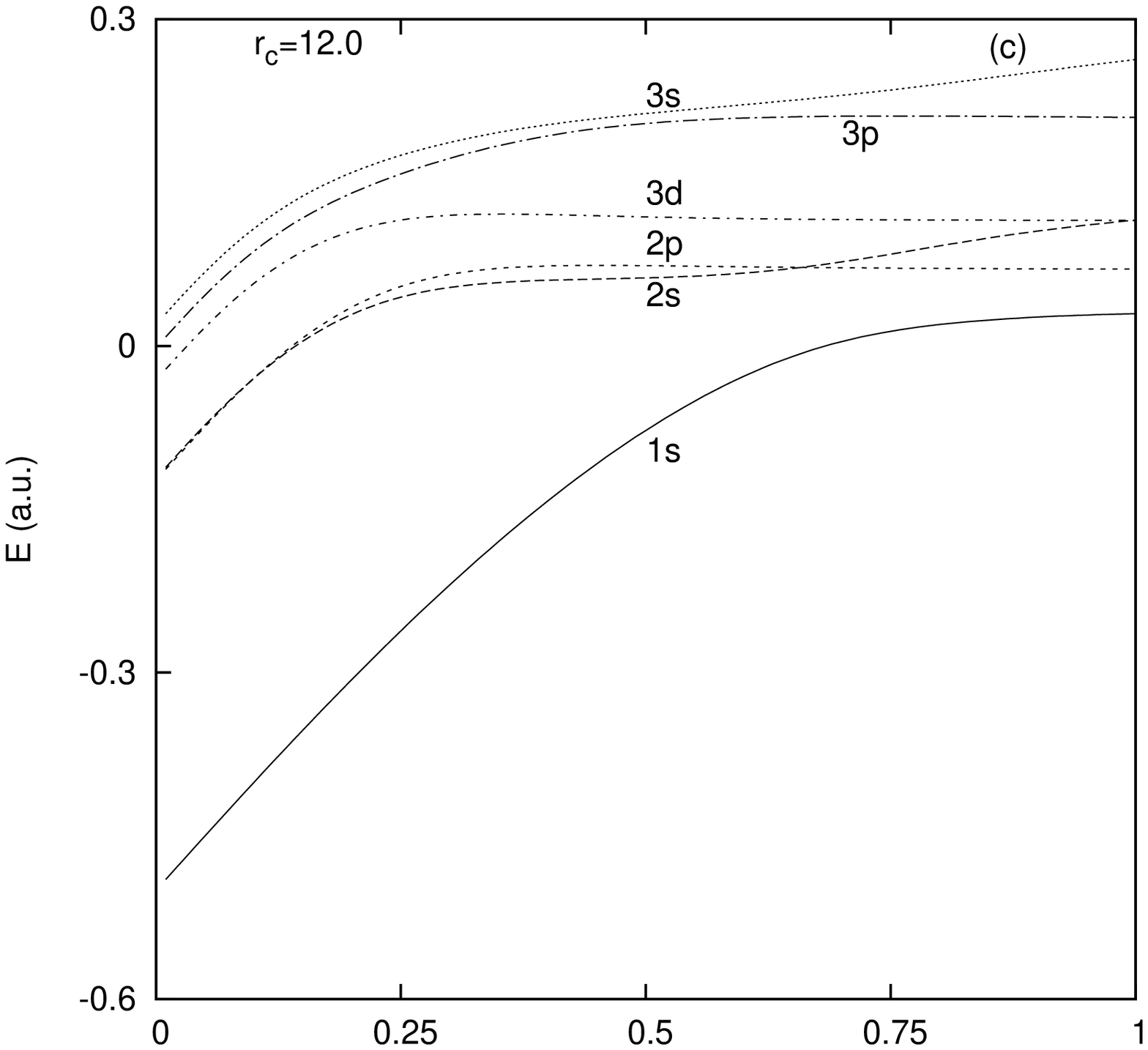}
\end{minipage}%
\begin{minipage}[c]{0.55\textwidth}
\centering
\includegraphics[scale=0.48]{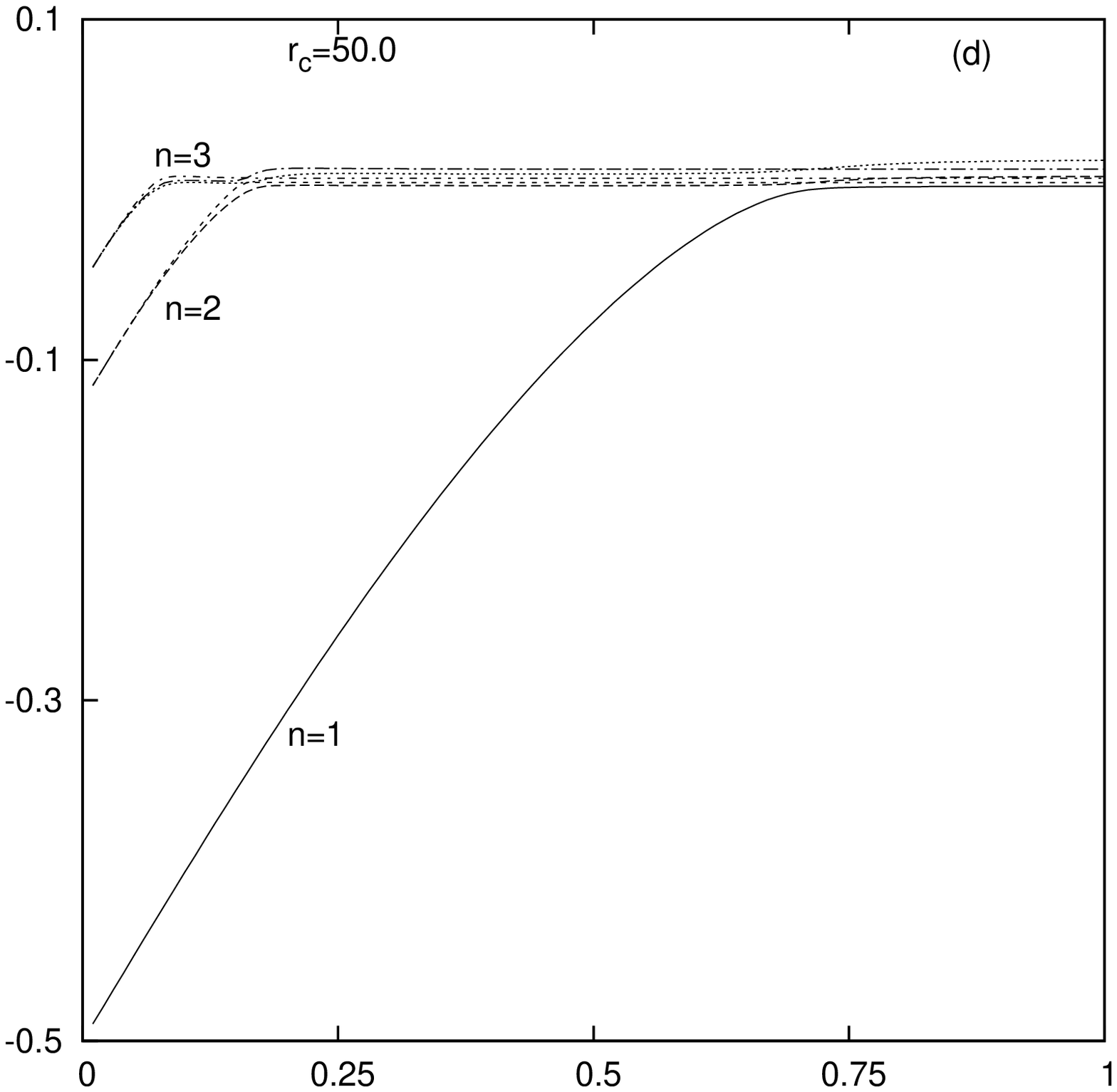}
\end{minipage}%
\\
\vspace{-1.4in}
\begin{minipage}[c]{0.55\textwidth}
\centering
\includegraphics[scale=0.48]{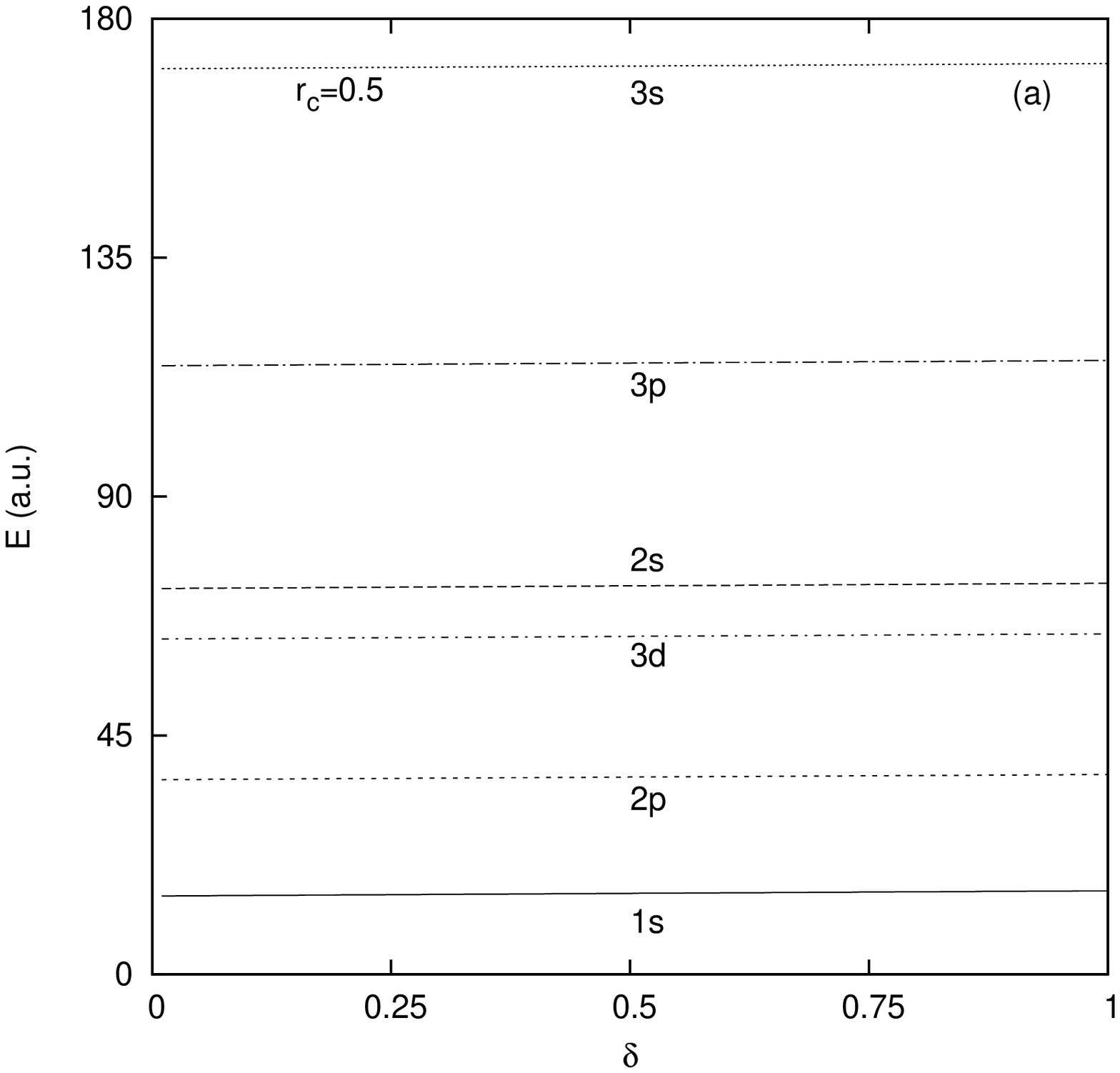}
\end{minipage}%
\begin{minipage}[c]{0.55\textwidth}
\centering
\includegraphics[scale=0.48]{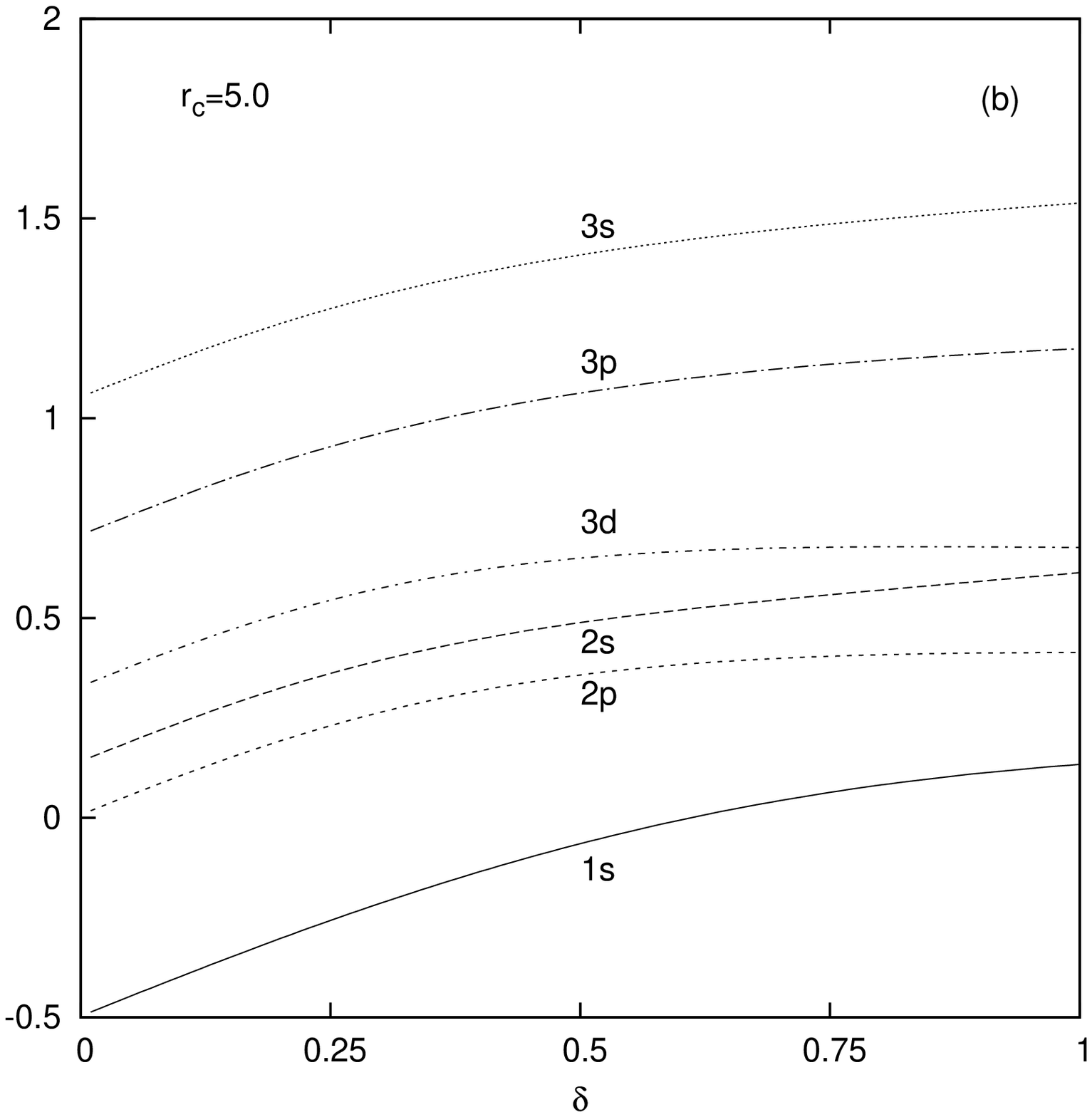}
\end{minipage}%
\vspace{-0.6in}
\caption[optional]{Energy changes (in a.u.) in confined ECSC potential with respect to $\delta$ for 6 states having 
$n=1,2,3$: (a)-(d) correspond to $r_c=0.5, 5, 12$ and 50 respectively. For more details, see text.}
\end{figure}

\begingroup
\squeezetable
\begin{table}
\caption {\label{tab:table3} Dipole polarizability (in a.u.) of confined ECSC potential with respect to 
cage radius.}
\begin{ruledtabular}
\begin{tabular}{c|ll|ll|ll|ll}
   $r_c$   & \multicolumn{4}{c}{$\delta=0.01$}    &  \multicolumn{4}{c}{$\delta=0.1$}   \\
\cline{2-9} 
       &  \multicolumn{2}{c}{$1s$}   &  \multicolumn{2}{c}{$2p$} &  \multicolumn{2}{c}{$1s$}  & \multicolumn{2}{c}{$2p$} \\
\cline{2-9} 
       &   $\alpha_D^K$  &  $\alpha_D^B$  &  $\alpha_D^K$        &   $\alpha_D^B$         
       &   $\alpha_D^K$  &  $\alpha_D^B$  &  $\alpha_D^K$        &   $\alpha_D^B$       \\
\cline{2-9} 
0.5   & 0.00200276  &  0.00203003   & 0.00378518  & 0.00420432  & 0.00200276  &  0.00203003 &  0.00378518  &  0.00420432 \\       
1     & 0.02847720  &  0.00286745   & 0.05871393  & 0.06487499  & 0.02847752  &  0.02867490 &  0.05871427  &  0.06487542 \\
2     & 0.34014205  &  0.34015656   & 0.87817541  & 0.95985795  & 0.34020627  &  0.34022090 &  0.87825964  &  0.95995987 \\
3     & 1.1732781   &  1.1809895    & 4.1219980   & 4.4548431   & 1.1743970   &  1.1820965  &  4.1240285   &  4.4572521  \\
4     & 2.2908131   &  2.3578294    & 11.964196   & 12.781332   & 2.2971117   &  2.3641982  &  11.983098   &  12.803289  \\
5     & 3.2040526   &  3.4029664    & 26.530803   & 28.013777   & 3.2214639   &  3.4215709  &  26.634577   &  28.131668  \\
8     & 3.9751111   &  4.4528509    & 119.98904   & 122.68212   & 4.0188523   &  4.5078558  &  123.13619   &  126.02491  \\
10    & 3.9986145   &  4.4968272    & 210.86134   & 212.42548   & 4.0449366   &  4.5567235  &  224.07520   &  225.96136  \\
15    & 4.0000555   &  4.5000719    & 369.95855   & 370.58964   & 4.0466744   &  4.5606489  &  454.93699   &  455.94767  \\
20    & 4.0000560   &  4.5000735    & 398.38484   & 400.95261   & 4.0466752   &  4.5606515  &  540.39106   &  548.31819  \\
30    & 4.0000560   &  4.5000735    & 400.17554   & 403.03939   & 4.0466752   &  4.5606515  &  558.54036   &  570.79672  \\
\end{tabular}
\end{ruledtabular}
\end{table}
\endgroup

We now discuss the effects of $\delta$ on energy changes in a spherically confined ECSC potential.
Figure 2 illustrates this for six states corresponding to $n=1,2,3$ at four selected $r_c$, namely, 
0.5, 5, 12 and 50 respectively in panels (a)-(d). The $\delta$ values are varied from 
0-1, while energy axis varies accordingly. For smaller confinement ($r_c=0.5$) as in (a), energies
change very small (practically remain unchanged) from their initial finite value, for the whole range of 
$\delta$. Individual plots are parallel and maintain good distance from each other, showing no 
mixing/crossing amongst the levels. With slight increase in box size (such as at $r_c=3$, which is not 
shown here), flat lines disappear giving rise to parallel curves. Energies drastically drop from (a) and 
relative separation between $2s, 3d$ levels reduces. Negative energy appears for ground state; however, 
energy sequence of (a) is maintained for all through out the $\delta$ range. Further increase in $r_c=5$ 
in (b), bends all the curves with still no mixing among them. But now $2p$ also comes closer in energy to 
those of $2s, 3d$ states. Further increase in $r_c=8$ (not shown here) leads to the onset of a general 
shape of these plots, which 
continues to remain unchanged for higher $r_c$'s as well. Initially in the smaller screening region, any 
increase in $\delta$ causes considerable increase in energy until reaching a certain threshold; after this 
energy changes with $\delta$ tends to be less dramatic. While all the levels maintain good distance from 
each other through out, $2s, 2p$ remain very close to each other for the most part of $\delta$; only starts 
to branch out at $\delta \approx 0.75$. Now as $r_c$ reaches 12 in (c), one clearly sees the three $n$'s 
making a family amongst each other. For smaller $\delta$, 
energy separation between $n=1$ and 2 are much larger compared to that between $n=2,3$. Once again 
$2s, 2p, 3d$ states continue to remain close to each other, especially for larger $\delta$. We also notice
$2s, 2p$ energies crossing at this stage. In the next $r_c$ (15 and 20, which are not given here), 
the set of plots corresponding to three $n$ continue to maintain their separate places; 
however evidently as $\delta$ increases more mixing amongst the states takes place and there is a tendency 
of all the six states to merge with increase in $r_c$. This is most conspicuously seen at a sufficiently
large $r_c=50$ in panel (d), where one effectively sees three plots corresponding to three $n$ quantum 
numbers. Thus, state $1s$ in $n=1$ remains a family of its own, while $2s, 2p$ forming another family 
practically merging at a much larger energy separation from $n=1$, in smaller $\delta$, and finally 
$3s,3p,3d$ coinciding with each other to form a third family. But $n=2$, 3 families join each other at 
about $\delta=0.25$, while $n=1$ merges with them at around $\delta=0.70$ so that after that all six states
continue to assume similar energy. Variations similar to those in Fig.~2 were constructed for nine states 
belonging to $n=4,5$ revealing many more mixing among states. These are not produced here to save space. 

At this stage, some specimen results are given for dipole polarizability of spherically confined ECSC potential
in Table~III. For this, we employ the simplified expressions given below, assuming that the original 
equations, derived for one-electron atoms in free-space, by Kirkwood and Buckingham, also hold good under 
confinement,
\begin{equation}
\alpha_D^K= \frac{4}{9} \langle r^2 \rangle^2; \ \ \ \ \ 
\alpha_D^B= \frac{2}{3} \left[ 
\frac{6 \langle r^2 \rangle^3 +3 \langle r^3 \rangle^2 - 8 \langle r \rangle \langle r^2 \rangle \langle r^3 \rangle} 
{9 \langle r^2 \rangle - 8 \langle r \rangle^2 }         \right]. 
\end{equation}
We restrict ourselves to $1s$, $2p$ states and $\delta=0.01$ and 0.1 in panels (a), (b) respectively. Thus 
$\alpha_D^K$ and $\alpha_D^B$ are offered for these two states at 11 selected values of $r_c$ covering a 
broad range. No results are reported for any of these. For a particular $\delta$, both $\alpha_D^K$ and 
$\alpha_D^B$ gradually increase with $r_c$, finally reaching an asymptotic value and also satisfying the 
bound $\alpha_D^K \leq \alpha_D^B$. Such an inequality is known to hold for \emph{free} spherically 
symmetric Coulomb potential. With increase in $r_c$, differences between the two $\alpha$ tend to increase 
significantly. {\color{red} Multipole (dipole, quadrupole and octupole) polarizabilities for free system 
has been reported \cite{lai13} through $B-$spline basis functions. Results of $\alpha_D^B$ for such 
systems at ten selected $\delta$ are compared with those. For weak to medium screening, the two results 
show agreement with each other. With increase in $\delta$, the disagreements set in.} One further notices that 
polarizabilities for $2p$ states are higher compared to the ground state for a given 
$\delta$; moreover, the asymptotic value for $2p$ state requires much higher $r_c$ compared to the ground
state. We also observe that, increase in $\delta$ causes both $\alpha_D^K$ and $\alpha_D^B$ to increase. 

\begingroup
\squeezetable
\begin{table}
\caption {\label{tab:table4} Comparison of {\color{blue} Buckingham} polarizabilies of ground state of \emph{free} ECSC potential, at 
selected $\delta$, with reference values of \cite{lai13}. PR implies Present Result. See text for details.}
\begin{ruledtabular}
\begin{tabular}{l|llllllllll}
 Set   & $\delta=0.0$   &  $\delta=0.05$   &  $\delta=0.1$   &  $\delta=0.15$   &  $\delta=0.2$   
       & $\delta=0.25$  &  $\delta=0.3$    &  $\delta=0.4$   &  $\delta=0.5$    &  $\delta=0.6$   \\  \hline
 PR    & 4.50000       & 4.50839         & 4.56065        &  4.68913         &  4.92553     
       & 5.3139        & 5.9289          & 8.5483         &  17.678          &  90.989          \\
 Ref.  & 4.50000       & 4.50839         & 4.56066        &  4.68918         &  4.92576  
       & 5.3147        & 5.9309          & 8.5607         &  17.746          &  91.310          \\

\end{tabular}
\end{ruledtabular}
\end{table}
\endgroup

Finally, Table~IV offers some results on $\alpha_D^B$ for the ground state of \emph{free} ECSC potential, at ten selected 
screening values. These are compared with the recent finite basis set calculation \cite{lai13} with $B$-spline functions, 
where dipole, quadrupole and octupole polarizabilities were reported. For mild to medium screening, present results show 
very good agreement with these reference values. As the screening parameter increases, disagreements start to build
up, which becomes significant for stronger screening. 

\section{conclusion}
Critical parameters in three screening potentials of physical interest have been investigated systematically. 
Accurate values of these are reported for Hulth\'en, Yukawa and ECSC potentials by means of GPS method; all the 
55 eigenstates on or below $n \leq 10$ are considered for all of them. Excellent agreement with best 
theoretical results are recorded. Some of these have not been published before. Additionally, 
we make an analysis of the energy spectrum in an ECSC potential when confined inside a spherically symmetric 
impenetrable wall, to simulate the H atom contained in a dense quantum plasma. Energy changes are followed 
with respect to cage radius and screening parameter. Besides, dipole polarizability and energy orderings are 
also discussed. 

\section{acknowledgment} The author thanks the three anonymous referees for their constructive comments. It is a pleasure 
to thank Drs.~Siladitya Jana and Sanjib Das for their kind help for figures. 
Final assistance from DST-SERB, New Delhi (Project No. EMR/2014/000838) is gratefully acknowledged.

\end{document}